\documentclass{ws-ijmpa-my}
\usepackage{amssymb}
\usepackage{amsmath}
\usepackage{graphicx}

\newcommand{\gd}{\delta}
\newcommand{\gl}{\lambda}
\newcommand{\gs}{\sigma}
\newcommand{\go}{\omega}

\newcommand{\rf}[1]{(\ref{#1})}

\begin{document}

\markboth{V.~S.~Rychkov}{Black hole production}

\catchline{}{}{}{}{}

\title{CLASSICAL BLACK HOLE PRODUCTION IN QUANTUM PARTICLE COLLISIONS}

\author{VYACHESLAV S.~RYCHKOV}
\address{Institute for
Theoretical Physics, University of Amsterdam,\\ 1018XE Amsterdam,
The Netherlands
%\\Email: rychkov@science.uva.nl
}

%%%%%%%%%%%%%%%%%%%%%%%%%%%%%%%%%%%%%%%%%%%%%%%%%%%%%%%%%%%%
% You may repeat \author \address as often as necessary    %
%%%%%%%%%%%%%%%%%%%%%%%%%%%%%%%%%%%%%%%%%%%%%%%%%%%%%%%%%%%%

\maketitle

%\begin{history}
%\received{DAY MONTH YEAR}
%\revised{DAY MONTH YEAR}
%\end{history}

\begin{abstract}
A semiclassical picture of black hole production in
trans-Planckian collisions is reviewed.
\end{abstract}


%%%%%%%%%%%%%%%%%%%%%%%%%%%%%%%%%%%%%%%%%%%%%%%%%%%%%%%%%%%%
% The main text of your paper   begins here              %
%%%%%%%%%%%%%%%%%%%%%%%%%%%%%%%%%%%%%%%%%%%%%%%%%%%%%%%%%%%%

\section{Introduction}
(This talk was based on Refs.~\refcite{bh1,R,GR}, where complete
bibliography can be found.)

 Let us compare two black hole (BH)
formation processes, happening on vastly different physical
scales. In one process, two solar mass BHs moving with
relativistic velocities collide to form a bigger BH. In the other,
a BH is formed in a collision of two trans-Planckian elementary
particles. ``Trans-Planckian" here means $E\gg 10^{19}$ GeV in
standard 4-dimensional gravity, or $E\gg 1$ TeV in Large Extra
Dimension scenarios of TeV-scale gravity.

The first, astrophysical, process is described by classical
General Relativity. What about the second one? In particular, what
is the role played by the quantum nature of elementary particles?
To make these questions more precise, let us first review the
classical gravity picture of BH production.
\section{Classical gravity picture}

\begin{figure}
\centerline{\psfig{file=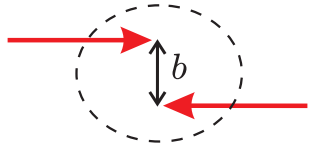,width=4cm}} \vspace*{8pt}
\caption{Grazing collision.}\label{grazing}
\end{figure}

\begin{figure}
\centerline{\psfig{file=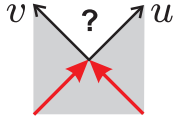,height=1.5cm}}
\vspace*{8pt} \caption{Collision spacetime of two Aichelburg-Sexl
waves (longitudinal slice).}\label{collision}
\end{figure}

\begin{figure}
\centerline{\psfig{file=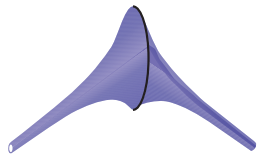,height=2cm}} \vspace*{8pt}
\caption{Closed trapped surface.}\label{CTS}
\end{figure}

In a totally classical description we consider a grazing collision
of two fast point particles of energy $E\gg 1$ in $D$-dimensional
Planck units (Fig. \ref{grazing}).
Naively, we expect a BH to form if the impact parameter $b$ is
comparable to the Schwarzschild radius of a $D$-dimensional BH of
mass $E$, where $D=4+n$, $n$ being the number of large extra
dimensions. In Planck units this corresponds to the condition:
\begin{align}\label{range}
    b\lesssim R_h \sim E^{1/(D-3)}\ .
\end{align}
This naive picture can be made precise by studying the geometry of
two colliding Aichelburg-Sexl shock waves, describing the fast
point particles. The colliding shocks have curvature concentrated
on the null planes $u=0$ and $v=0$ (Fig.~\ref{collision}).
Spacetime is flat before and after the shocks. The interaction
region $u,v>0$ will be curved, and the metric there is unknown. To
prove the BH formation, one can look for a closed trapped surface
(CTS, also called apparent horizon) in the known part of
spacetime. In Refs.~\refcite{EG,YN} such apparent horizons were
indeed found for impact parameters in the range \rf{range}. They
have a rather peculiar shape, consisting of two throats narrowing
along the world lines of colliding particles and glued together at
the transverse collision plane $u=v=0$ (Fig.~\ref{CTS}).

\section{Why classical gravity is applicable}

Two questions can be raised concerning the applicability of
classical gravity in this problem:
\begin{itemize}
    \item We are dealing with a potentially violent process occurring at Planckian
    energies. Shouldn't quantum gravity effects become significant?
    \item In accelerators, colliding particles are definitely not pointlike. Instead, they come
    in wide wavepackets of macroscopic size. Isn't this in direct
    contradiction with the point-particle approximation used in the
    above argument?
\end{itemize}
As it will turn out, in a more careful treatment these potential
problems in a sense compensate each other, and the final result is
not affected.

\subsection{Wavepacket subdivision}

The point-particle approximation used above cannot be taken over
to quantum theory. Indeed, quantum description should be
compatible with the usual position uncertainty, which for an
ultrarelativistic particle is of the order of the wavelength
$E^{-1}$. To incorporate this uncertainty we should consider
wavepackets of finite spread $w\gtrsim E^{-1}$.

In fact, it is quite obvious that the CTS argument does not need
particles to be exactly point-like, and works under a much milder
assumption of them having size $w\ll R_h$. The resulting collision
spacetime will be a small perturbation of the point-particle
spacetime and will still contain a CTS (\cite{GR}).

\begin{figure}
\centerline{\psfig{file=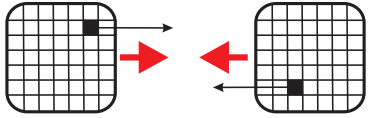, height=1.5cm}}
\vspace*{8pt}
 \caption{Subdivided wavepackets} \label{wavepackets}
\end{figure}

However, the wavepackets describing particles in a collider beam
have much larger, macroscopic size set by the beam radius ($\sim
10^{-3}$ cm at the
LHC). %(16.7 micrometers to be precise, Particle Data Group)
To study BH production, these huge wavepackets should thus be
subdivided into smaller wavepackets of size $w$ satisfying the
previous two limits:
\begin{equation}\label{limits}
    E^{-1}\lesssim w \ll R_h
\end{equation}
This subdivision (see Fig.~\ref{wavepackets})
can be carried out in such a manner that different small
wavepackets correspond to almost orthogonal states. (This
orthogonality is obvious in position space, although it would be
hard to see in the momentum space.) Because of this orthogonality,
to compute the total BH production cross section, we must simply
count the total number of pairs of small wavepackets, for which a
BH forms (this is a 0-1 possibility). It is easy to see that such
counting results in the geometric cross section.

By this argument we reduced the problem to analyzing collision of
two wavepackets of size $w$. We will now show that there exist $w$
satisfying \rf{limits}, for which quantum gravity corrections to
the semiclassical description are small.

\subsection{Curvature test}
\begin{figure}
\centerline{\psfig{file=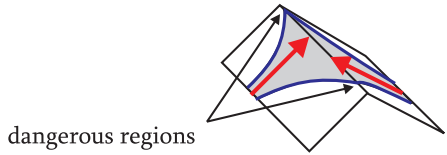,height=2cm}} \vspace*{8pt}
\caption{A section of CTS from Fig.~\ref{CTS}. The gray area will
eventually become the BH interior.}\label{CTSsection}
\end{figure}

 For the
classical solution to be stable with respect to quantum
corrections, curvature must be small ($\ll 1$ in Planck units).
The Riemann tensor of the Aichelburg-Sexl shock wave corresponding
to the left-moving particle has $\delta$-function components
($x_i$ are $D-2$ transverse coordinates, $r=|x|$):
\begin{equation}
\label{Riemann} R_{uiuj}\propto E\,\delta(u)\,\partial_i\partial
_j(1/r^{D-4})\ .
\end{equation}
This field should be superposed with the similar field of the
right-moving particle, shifted by $b$ in the transverse direction.
This right-moving shock wave will have large $R_{vivj}$
components, and we can form a nonvanishing curvature invariant:
\begin{equation}
\label{Rsq} (R_{\mu\nu\gl\gs})^2 \sim E^{-\frac
2{D-3}}\,\gd(u)\,\gd(v)\qquad (r\sim R_h).
\end{equation}
This estimate would seem to imply that curvature blows up on the
CTS at the points where it crosses the transverse collision plane
(see Fig.~\ref{CTSsection}). However, this is where the wavepacket
width comes to the rescue. To take it into account, we have to
smear out $\delta$-functions in \rf{Rsq} on scale $w$. To keep
integral equal 1, the effective maximal value of $\gd(u)$,
$\gd(v)$ becomes $\sim 1/w$. Thus the final curvature estimate in
the vicinity of the CTS takes the form \cite{bh1,R,GR}:
\begin{equation}\label{width}
(R_{\mu\nu\gl\gs})^2 \sim R_h^{-2} w^{-2}\qquad (r\sim R_h).
\end{equation}
From this we see that curvature can be kept small as long as
    $w\gg R_h^{-1}$ (\cite{GR}),
a condition compatible with the allowed range \rf{limits}.

\subsection{Graviton counting test}
\label{count} Another necessary condition for the classical
gravity description to be valid is that quantum fluctuations of
the gravitational field have to be be suppressed compared to its
classical value. This will happen when graviton occupation numbers
of the colliding fields are large. The concept of gravitons should
be applicable for linearized gravity, when deviation from the
Minkowski metric is small:
\begin{equation}\label{linear}
    g_{\mu\nu}=\eta_{\mu\nu}+h_{\mu\nu},
    \qquad\left|h_{\mu\nu}\right|\ll 1.
\end{equation}
In this case $h_{\mu\nu}$ can be quantized as a free field, with
quanta being transverse gravitons. In our case this condition
turns out to be satisfied near the shock fronts.

\begin{figure}
\centerline{\psfig{file=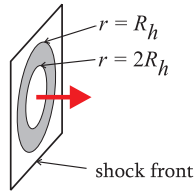,height=2.5cm}}
\vspace*{8pt} \caption{Gravitons in the gray area need to be
counted.}\label{shockfront}
\end{figure}

\begin{figure}
\centerline{\psfig{file=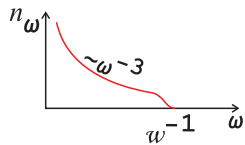,height=1.8cm}} \vspace*{8pt}
\caption{Gravitons spectrum in the shock front.}\label{spectrum}
\end{figure}

Thus, we would like to count gravitons contained in the
gravitational field of colliding particles, say, of the
right-moving one. The presence of the other particle does not play
a role in this counting before the shock waves collide. More
specifically, we would like to count gravitons contained in the
shock front at $r\sim R_h$, since this is the region relevant for
the apparent horizon formation (see Fig.~\ref{shockfront}).
A detailed calculation (\cite{R}) shows that the spectrum of
gravitons is given by
\begin{align}\label{}
    n_\go \sim R_h^{D-4}\omega^{-3}.
\end{align}
This formula is valid up to frequencies $\go\lesssim w^{-1}$, at
which point the spectrum cuts off (see Fig.~\ref{spectrum}).
The total number of gravitons with energy $\sim \Omega$ is
\begin{equation}
    \label{Ntot}
    N_{\Omega}\sim \int_{\omega\sim \Omega} d\omega\,n_\omega\sim
    R_h^{D-4}\Omega^{-2}.
\end{equation}
Condition $N_\Omega\gg 1$ becomes most restrictive when applied at
$\Omega\sim w^{-1}$, which is the smallest scale present in the
classical solution. This implies that $w$ has to satisfy $w\gg
R_h^{2-D/2}$ (\cite{GR}). Again we see that this is compatible
with \rf{limits}.

\section{Conclusions}

In this talk we argued that it \emph{is} possible to carry out the
analysis of BH production in trans-Planckian elementary particle
collisions in controlled semiclassical approximation. We showed
that both criteria of semiclassicality (low curvatures, high
quantum numbers) are satisfied. The geometric cross section
estimate is in good shape.

%\section*{Acknowledgements}

I am grateful to Steve Giddings for an opportunity to collaborate
on justifying the semiclassical approximation of the BH production
process. This research was supported by Stichting FOM.

%%%%%%%%%%%%%%%%%%%%%%%%%%%%%%%%%%%%%%%%%%%%%%%%%%%%%%%%%%%%
% Doing Appendix(ices) - Appendix A & B are shown below    %
%%%%%%%%%%%%%%%%%%%%%%%%%%%%%%%%%%%%%%%%%%%%%%%%%%%%%%%%%%%%

%%%%%%%%%%%%%%%%%%%%%%%%%%%%%%%%%%%%%%%%%%%%%%%%%%%%%%%%%%%%
% Doing references:                                %
%%%%%%%%%%%%%%%%%%%%%%%%%%%%%%%%%%%%%%%%%%%%%%%%%%%%%%%%%%%%

\end{document}